# On the Convergence of the Electronic Structure Properties of the FCC Americium (001) Surface

Da Gao and Asok K. Ray*
Department of Physics
P. O. Box 19059
University of Texas at Arlington
Arlington, Texas 76019

*akr@uta.edu.

# Abstract

Electronic and magnetic properties of the fcc Americium (001) surface have been investigated via full-potential all-electron density-functional electronic structure calculations at both scalar and fully relativistic levels. Effects of various theoretical approximations on the fcc Am (001) surface properties have been thoroughly examined. The ground state of fcc Am (001) surface is found to be anti-ferromagnetic with spin-orbit coupling included (AFM-SO). At the ground state, the magnetic moment of fcc Am (001) surface is predicted to be zero. Our current study predicts the semi-infinite surface energy and the work function for fcc Am (001) surface at the ground state to be approximately 0.82 $J/m^2$ and 2.93 eV respectively. In addition, the quantum size effects of surface energy and work function on the fcc Am (001) surface have been examined up to 7 layers at various theoretical levels. Results indicate that a three layer film surface model may be sufficient for future atomic and molecular adsorption studies on the fcc Am (001) surface, if the primary quantity of interest is the chemisorption energy.

## I. Introduction

Actinides, the elements from actinium and beyond in the periodic table including uranium, plutonium, americium and other radioactive elements, are well-known nuclear weapon materials. In as much as they are praised for their role in ending the Cold War, understandable concerns persist about their potentially catastrophic impact on the environment and global health [1]. On the scientific front, the miraculous *5f* electron properties of the actinides, being strongly correlated and heavy fermion systems, especially their bonding properties, are still not well understood because of their inherent complexities. To address these issues, a significant amount of theoretical study is required since experimental work is relatively difficult to perform on actinides due to material problems and toxicity. Moreover, such studies are essential for the actinides' rational use, safe disposition, and reliable long-term storage.

In the actinides, americium occupies a pivotal position with regard to the *5f* electron properties [2]. First, although the atomic volume of the actinides before Pu continuously decreases as a function of the increasing atomic number from Ac until Np, a sharp atomic volume increase from Pu to Am has been observed [3]. Such observation indicates that the *5f* electron properties have changed dramatically starting from somewhere between Pu and Am. It is believed that the *5f* electrons of the actinides before Am are delocalized and participate in bonding while the *5f* electrons of the actinides after Pu become localized and non bonding [4, 5]. Our earlier fully relativistic density functional studies of the fcc Am bulk and the Am monolayer properties, using both local density and generalized gradient approximations, as well as other published works in the literature have supported the localized picture for Am [6-8]. Second, as the applied pressure is increased, the americium metal displays different crystal structures [9]: double hexagonal close packed (Am I) → face-centered cubic (Am II) → face-centered orthorhombic (Am III) → primitive orthorhombic (Am IV). Such behaviors could further provide more insights of the americium 5*f* electron properties, especially the Mott transition, *i.e.*, the evolution of the 5*f* electrons from localized to the delocalized. The experimental studies have shown that the 5*f* electrons of Am III and Am IV are probably delocalized [2, 9]. However, recent density-functional electronic structure calculations with respect to the high-pressure behavior of americium found only the fourth phase (Am IV) to be delocalized and the 5*f* electrons of the three previous americium phases to be primarily localized [10]. Such controversies clearly indicate that further experimental and theoretical studies are required to improve our understanding of americium and the associated 5*f* electrons. Third, the electronic structure of americium metal, whereby six *f* electrons form an inert core, decoupled from the *spd* electrons that control the physical properties of the material, have contributed to superconductivity in Am [11, 12]. A recent high-pressure study of the resistivity of americium metal has been reported and deduced the unusual dependence of the superconducting temperature on pressure [13].

However, to the best of our knowledge, *very few* studies exist in the literature about the Am surface despite the fact that such surface studies may lead to a better understanding of the detailed americium surface corrosion mechanisms in the presence of environmental gases and thus help addressing the environmental consequences of nuclear materials. Moreover, surface studies may provide an effective way to probe the americium 5*f* electron properties and their roles in the chemical bonding. The unusual aspects of the bonding in bulk americium are apt to be enhanced at a surface or in a thin layer of americium adsorbed on a substrate as a result of the reduced atomic coordination of a surface atom and the narrow bandwidth of surface states. Thus, americium surfaces and thin films may also provide valuable information about the bonding in americium. Recently, Gouder *et al.* have studied thin films of Am, AmN, AmSb, and $Am_2O_3$ prepared by sputter deposition by x-ray and ultraviolet photoelectron spectroscopy (XPS and UPS) techniques. Their experimental studies indicate that in all four Am systems, *5f* electrons are largely localized, though the XPS core-level spectrum of Am indicates residual *5f* hybridization [14].

We have recently investigated, in detail, the surface properties of fcc δ–Pu and atomic and molecular adsorptions on such surfaces [15]. At the same time, we are also particularly intrigued by the surface properties of Am, the nearest neighbor of Pu. Both Pu and Am represent the boundary position between the light actinides, Th to Pu, and the heavy actinides, Am and beyond. Pu has an open shell of *f* electrons while Am is closer to a full $j = 5/2$ shell. In addition, the transition from the itinerant to localized *5f* electrons takes place somewhere between Pu and Am; yet there is no such apparent transition observed, at least, in α-Pu although the *5f* electrons of δ-Pu are partially localized [16], as indicated by its atomic volume, which is approximately halfway between α-Pu and Am. Thus we believe systematic and fully relativistic density functional studies of Pu and Am surface chemistry and physics could certainly lead to more insights and knowledge about the actinides.

In this paper, for the first time we report on the fcc Am (001) surface properties and a detailed comparison with the published fcc δ-Pu (001) surface properties. For such surface studies, it is common practice to model the surface of a semi-infinite solid with an ultra thin film (UTF), which is thin enough to be treated with high-precision density-functional calculations, but is thick enough to realistically model the intended surface. Determination of an appropriate UTF thickness is complicated by the existence of possible quantum oscillations in UTF properties as a function of thickness, the so-called quantum size effect (QSE). These oscillations were first predicted by calculations on jellium films [17, 18] and were subsequently confirmed by band-structure calculations on free-standing UTFs composed of discrete atoms [19-22]. The adequacy of the UTF approximation obviously depends on the size of any QSE in the relevant properties of the model film. Thus, it is important to determine the magnitude of the QSE in a given UTF prior to using that UTF as a model for the surface. This is particularly important for Am films, since the strength of the QSE is expected to increase with the number of valence electrons [17]. Consequently, this study has also examined the QSE of the fcc Am (001) surface.

## II. Computations

The present calculations have been carried out using the full-potential all-electron method with a mixed basis set of linearized-augmented-plane-wave (LAPW) and augmented-plane-wave plus local orbitals (APW+lo), with and without spin-orbit coupling (SO), as implemented in the WIEN2K suite of programs [23-25]. Six theoretical levels of approximation, *i.e.,* anti-ferromagnetic-no-spin-orbit-coupling (AFM-NSO), anti-ferromagnetic-spin-orbit-coupling (AFM-SO), spin-polarized-no-spin-orbit-coupling (SP-NSO), spin-polarized-spin-orbit-coupling (SP-SO), non-spin-polarized-no-spin-orbit-coupling (NSP-NSO), and non-spin-polarized-spin-orbit-coupling (NSP-SO) have been implemented in our calculations. The generalized-gradient-approximation (GGA) to density functional theory [26] with a gradient corrected Perdew- Berke - Ernzerhof (PBE) exchange-correlation functional [27] is used and the Brillouin-zone integrations are conducted by an improved tetrahedron method of Blöchl-Jepsen-Andersen [28]. In the WIEN2k code, the alternative basis set APW+lo is used inside the atomic spheres for the chemically important orbitals that are difficult to converge, whereas LAPW is used for others. The local orbitals scheme leads to significantly smaller basis sets and the corresponding reductions in computing time, given that the overall scaling of LAPW and APW + lo is given by $N^3$, where N is the number of atoms. In addition, results obtained with the APW + lo basis set converge much faster and often more systematically towards the final value [29]. As far as relativistic effects are concerned, core states are treated fully relativistically in WIEN2k and for valence states, two levels of treatments are implemented: (1) a scalar relativistic scheme that describes the main contraction or expansion of various orbitals due to the mass-velocity correction and the Darwin s-shift [30] and (2) a fully relativistic scheme with spin-orbit coupling included in a second variational treatment using the scalar-relativistic eigenfunctions as basis [31]. The present computations have been carried out at both scalar-relativistic and fully-relativistic levels to determine the effects of relativity. To calculate the total energy, a constant muffin-tin radius ($R_{mt}$) of 2.60 a.u. is used for all calculations. The plane-wave cut-off $K_{cut}$ is determined by $R_{mt} K_{cut} = 9.0$. The (001) surface of fcc Am is modeled by periodically repeated slabs of N Am layers (with one atom per layer and N=1-7) separated by an 80 a. u. vacuum gap. We have tested different vacuum gap lengths in order to determine an appropriate length such that an accurate result can be achieved without an excessive increase in computational costs. Twenty-one irreducible K points have been used for reciprocal-space integrations. For each calculation, the energy convergence criterion is set to be 0.01 mRy. Due to severe demands on computational resources due to all-electron calculations and internal consistency, we have chosen the calculated equilibrium lattice constants, obtained at different levels of approximation for bulk fcc Am [6], also in the surface computations at the corresponding level of approximation. No further relaxations and/or reconstructions have been taken into account. It is not expected that this will change the primary conclusions of this paper.

**III. Results and Discussions**

We first calculated the total energies for fcc Am (001) films at all six theoretical levels, namely NSP-NSO, NSP-SO, SP-NSO, SP-SO, AFM-NSO, and AFM-SO levels, and plotted the results in Fig. 1, respectively. For comparison, the total energies of the fcc Am bulk at the corresponding theoretical level are also plotted in Fig. 1. Our results showed that with spin orbit coupling, the total energies of fcc Am (001) films are much lower, about 0.60 Ry/atom, than those without spin orbit coupling, as has been observed also before in the bulk properties. A similar spin polarization effect on the total energies of Am (001) films is found here with or without the inclusion of spin orbit coupling effects. From the total energy point of view, the Am (001) films at the AFM-SO level have the lowest total energy and therefore the AFM-SO is the ground state of the Am (001) films, which is consistent with the former theoretically predicted fcc Am bulk ground state [6, 10, 32], though the ground state of Am is experimentally believed to be nonmagnetic [8]. In our previous electronic structure studies [15] of bulk δ-Pu and the δ-Pu surfaces, the ground state is also found to be AFM-SO.

We also calculated the cohesive energies $E_{coh}$ for the fcc Am (001) N-layer films with respect to N monolayers and found that the cohesive energy increases monotonously with the film thickness (shown in Fig. 2 and Table I) at all six levels of calculations. It is also observed that the rate of increase of cohesive energy drops significantly as the number of layers increases, which has been previously noticed for the cohesive energy of δ-Pu (001) surface as well, and we expect that the convergence in the cohesive energy can be achieved after a few more layers. However, since to the best of our knowledge, the experimental value for the semi-infinite surface cohesive energy is not known, we are unable to predict how many layers will be needed to achieve the semi-infinite surface energy. Typically, spin polarization lowers the cohesive energy by ~32%-50% at both the scalar relativistic and fully relativistic levels of theory. On the other hand, spin-orbit coupling increases the cohesive energy of the spin-polarized N-layers by about ~11-13% but reduces the cohesive energy of the non-spin-polarized N-layers by about ~9-15%. These features are in general agreement with the results of δ-Pu (001) surface [15]. At the antiferromagnetic state, spin-orbit coupling increases the cohesive energy of the N-layers by about ~13-15%. All cohesive energies are positive, indicating that all layers of Am (001) films are bound relative to the monolayer.

The incremental energies $E_{inc}$ of N layers with respect to (N-1) layers plus a single monolayer are also calculated and plotted (shown in Fig. 3 and Table I). The incremental energies are found to become relatively stable when the number of layers is greater than five. The study of the δ-Pu (001) films shows the incremental energies quickly saturated once the number of δ-Pu (001) layers is greater than three [15]. We also notice that both $E_{coh}$ and $E_{inc}$ of fcc Am (001) films are smaller than those of δ-Pu (001) films. Such difference indicates that the layers of δ-Pu (001) films are strongly bound relative to the monolayer, compared to layers of fcc Am (001) films. This is attributed to the *5f* electron properties of fcc Am and fcc δ-Pu, namely, that the *5f* electrons are more localized in fcc Am rather than in δ-Pu.

To further address the effects brought by the spin polarization and the spin-orbit coupling, we also calculated spin-polarization energies and spin-orbit coupling energies for the Am (001) films at various theoretical levels, and the results are shown in Table II as well as in Fig. 4 and Fig. 5. The spin-polarization energy $E_{sp}$ is defined by

$$E_{SP} = E_{tot}(NSP) - E_{tot}(SP), \qquad (1)$$

and the spin-orbit coupling energy $E_{so}$ is defined by

$$E_{so} = E_{tot}(NSO) - E_{tot}(SO), \qquad (2)$$

We note that both spin polarization and spin orbit coupling energies become rather stable when the number of layers equals three. It can also be seen that spin-orbit coupling plays a more important role than spin-polarization in reducing the total energies of the fcc Am (001) films, *i.e.*, spin-orbit coupling effect reduces the total energy by 7.56-8.94 eV/atom, while spin-polarization effect decreases the total energy only by 1.52-3.53 eV/atom. Comparing these to the SO coupling and SP effects in the δ-Pu (001) films, which are 7.11-7.99 eV/atom and 0.4-1.9 eV/atom respectively [14], the effects in fcc Am (001) films are more pronounced, especially the spin polarization effect. The discrepancy can also be partially attributed to the additional *5f* electron in Am and the localized feature of these electrons.

As far as the magnetic properties of fcc Am (001) surface are concerned, the spin magnetic moments of fcc Am (001) films at the SP-NSO, SP-SO, AFM-NSO, and AFM-

SO levels have been calculated and the results are listed in Table II. We also plotted these magnetic moments for Am (001) films as a function of the number of Am layers in Fig.6. Several features have been observed for the magnetic properties. First, for the Am (001) films at both AFM-NSO and AFM-SO levels, the magnetic moments show a behavior of oscillation, which becomes smaller with the increase of the number of layers, and gradually the magnetic moments approach zero. The Am (001) films with an odd number of layers have magnetic moments decreasing with the increase of the number of layers, while the Am (001) films with an even number of layers always have zero magnetic moments. Second, for the Am (001) films at the SP-NSO and SP-SO levels, the magnetic moments are, in general, larger than the corresponding bulk values of 7.32 and 6.90 $\mu_B$/atom [6], and with the increase of the number of layers the magnetic moments quickly approach the values of the corresponding bulks. For the seven layers thick film, the magnetic moment at the SP-NSO and SP-SO levels is 7.32 and 6.95 $\mu_B$/atom already. The spin magnetic moments of δ-Pu (001) films at the SP-NSO and SP-SO levels show a similar feature as found here except that the magnetic moments of δ-Pu (001) films are smaller than the corresponding magnetic moments of Am (001) films. The difference is attributed to the additional *5f* electron in Am. Third, for the Am (001) films at the anti-ferromagnetic state, spin-orbit coupling has negligible effects on the magnetic properties while for the Am (001) films at the spin-polarized states, spin-orbit coupling lowers the magnetic moments about 0.37 $\mu_B$/atom.

We also studied the quantum size effects in the fcc Am (001) films. The relevant physical quantities if interest here are typically the surface energies and work functions. We have calculated the surface energy for a *N*-layer film from [33]

$$E_s = (1/2)[E_{tot}(N) - NE_B], \qquad (3)$$

where $E_{tot}(N)$ is the total energy of the $N$-layer slab and $E_B$ is the energy of the infinite crystal. Assuming $N$ is large enough and $E_{tot}(N)$ and $E_B$ are known to infinite precision, then Eq. (3) is exact. However, if the bulk and film calculations are not entirely consistent with each other, then $E_s$ will linearly diverge with increasing $N$. Stable and internally consistent estimates of $E_s$ and $E_B$ can, however, be extracted from a series of values of $E_{tot}(N)$ via a linear least-squares fit to [34]

$$E_{tot}(N) = E_B N + 2E_s, \qquad (4)$$

To obtain an optimal result, the fit to Eq. (4) should only be applied to films which include, at least, one bulk-like layer, *i.e.*, $N > 2$. This fitting procedure has been independently applied to the fcc Am (001) films at all six levels of theory, respectively. Accordingly, six values of $E_B$, i.e., -61041.70707, -61042.35104, -61041.89126, -61042.45788, -61041.89118, and -61042.46001 Ry, and six values of semi-infinite surface energy $E_s$, i.e., 1.56, 1.48, 0.69, 0.80, 0.73, and 0.82 J/m$^2$, are derived for the Am (001) films at the NSP-NSO, NSP-SO, SP-NSO, SP-SO, AFM-NSO, and AFM-SO levels, respectively. We note that the semi-infinite surface energy decreases by about forty seven percent from the NSP-NSO level to the AFM-SO level. The surface energy for each film has been computed using the calculated $N$-layer total energy and appropriate fitted bulk energy. We have listed the results in Table III and also plotted the predicted surface energies as a function of the number of Am layers in Fig. 7. Several features of the fcc Am (001) surface energies are notable from our results. First, for all the theoretical levels except the NSP-SO level, the surface energy of Am (001) films converges pretty well to the corresponding semi-infinite surface energy when the number of layers reaches three. A similar behavior is also found in the δ-Pu (001) films [15]. From these results, we infer that a three layer film surface model may be sufficient for future atomic and molecular adsorption studies on the Am (001) surface, if the primary quantity of interest is the chemisorption energy. Second, our results show that spin polarization plays a more significant effect on the Am (001) surface energy than the spin orbit coupling does. To be specific, spin polarization lowers the Am (001) surface energy from ~1.5 J/m$^2$ down to ~0.8 J/m$^2$. Third, the surface energy of Am (001) films at the ground state, i.e., AFM-SO level, is predicted to be about 0.82 J/m$^2$.

The work function $W$ of the Am (001) surface is calculated according to the following formula

$$W = V_0 - E_F, \tag{5}$$

where $V_0$ is the Coulomb potential energy at the half height of the slab including the vacuum layer and $E_F$ is the Fermi energy. We have calculated the work functions of Am (001) films up to seven layers at all six theoretical levels, i.e., NSP-NSO, NSP-SO, SP-NSO, SP-SO, AFM-NSO, and AFM-SO, respectively. The results are listed in Table III and plotted in Fig. 8 as well. It can be observed that the work functions show some oscillations at all six theoretical levels up to seven layers. This is different from the δ-Pu (001) films where no significant work function oscillation was observed beyond five layers [15]. Furthermore, no clear even-odd oscillatory pattern is shown here. The current results indicate that at least a 7-layer film is necessary for any future adsorption investigation that requires an adsorbate-induced work function shift. The work functions of Am (001) films at seven layers are calculated to be 3.30, 3.28, 2.67, 2.78, 2.78, and

2.93 eV at the NSP-NSO, NSP-SO, SP-NSO, SP-SO, AFM-NSO, and AFM-SO levels, respectively. These values are, in general, smaller than the work functions of δ-Pu (001) films [15].

The 5*f* electrons in the fcc Am bulk are found to be primarily localized [2, 6, 9, 10]. However, the phase transition from fcc Am II to Am III is still controversial [2, 9, 10], as mentioned earlier. We thus explored the density of states (DOS) of 5*f* electrons in fcc Am (001) films at the ground state (AFM-SO level) with various number of layers (N=1, 4, 7) in order to learn how the 5*f* electron properties in the Am (001) thin films change as a function of the number of layers. The results have been plotted in Fig. 9. From the figure, we first note that the two 5*f* peaks, one below the Fermi level while the other above the Fermi level, are well separated by a wide gap indicating that the 5*f* electrons are localized. The gap width is about 2 eV for the present Am (001) surface calculations, which is in good agreement with the gap width found in the bulk dhcp Am [10] and bulk fcc Am [6]. In addition, as the thickness of the Am (001) thin films increases, the center of the first peak appears to be moving away from the Fermi level, signifying more localized 5*f* electrons, which is consistent with a recent photoemission study of the electronic structure of pure Am and Am compounds films [14]. In contrast to this, the thickness dependence of the degree of 5*f* electron delocalization in δ-Pu (001) films is varying [15]. Our current results indicate that as Am (001) films become thicker, the 5*f* electrons in the Am (001) films tend to behave more like the 5f electrons in the bulk phase. On the contrary, the 5*f* electrons are much less localized in Am (001) monolayer than in the bulk. This means more 5*f* electrons participate in chemical bonding in Am (001) monolayer, implying that the relaxed lattice constants in Am (001) monolayer should be smaller than that in Am bulk. Our previous study of the Am and Pu monolayer properties [6, 35] have shown a compression phenomena of the monolayers, which is in agreement with the present work.

## IV. Conclusions

We have performed full-potential all-electron density-functional electronic structure study of the fcc Am ultrathin (001) films at both scalar and fully relativistic levels. Our present calculation provides the first electronic structure results for fcc Am (001) surface. In order to examine effects of various theoretical approximations including spin-polarization and spin-orbit coupling, our current study has been carried out at six theoretical levels, namely, NSP-NSO, NSP-SO, SP-NSO, SP-SO, AFM-NSO, and AFM-SO. Our results show that the ground state of fcc Am (001) surface is the anti-ferromagnetic state with spin-orbit coupling included. At the ground state, the magnetic moment of fcc Am (001) surface is zero, which is in good agreement with the zero magnetic moment found in the ground state Am metal [6, 8]. Our results indicate that spin polarization lowers the cohesive energy by ~32%-50% at both the scalar relativistic and fully relativistic levels of theory. On the other hand, spin-orbit coupling increases the cohesive energy of the spin-polarized N-layers by about ~11-13% but reduces the cohesive energy of the non-spin-polarized N-layers by about ~9-15%. At the antiferromagnetic state, spin-orbit coupling increases the cohesive energy of the N-layers (001) Am films by about ~13-15%.

Our present results predict the semi-infinite surface energy and the work function for fcc Am (001) surface at the ground state to be approximately 0.82 J/m$^2$ and 2.93 eV respectively. The corresponding quantum size effects on the fcc Am (001) films have also been investigated up to 7 layers at different theoretical levels. We found that at least a 7-layer film surface model is necessary for any future fcc Am (001) surface adsorption investigation that requires an adsorbate-induced work function shift. Fnally, the 5$f$ electrons in fcc Am (001) thin films are found to be more localized as the films become thicker. In contrast to this, the thickness dependence of the degree of 5$f$ electron delocalization in δ-Pu (001) films is varying.

**Acknowledgments**

This work is supported by the Chemical Sciences, Geosciences and Biosciences Division, Office of Basic Energy Sciences, Office of Science, U. S. Department of Energy (Grant No. DE-FG02-03ER15409) and the Welch Foundation, Houston, Texas (Grant No. Y-1525).

Table I. Cohesive energies $E_{coh}$ per atom with respect to the monolayer and incremental energies $E_{inc}$ for fcc Am (001) N layers (N = 1-7) at all six theoretical levels.

| N | Theory | $E_{coh}$ (eV/atom) | $E_{inc}$ (eV) |
|---|---|---|---|
| 2 | NSP-NSO | 1.37 | 2.75 |
|   | NSP-SO  | 1.16 | 2.32 |
|   | SP-NSO  | 0.69 | 1.39 |
|   | SP-SO   | 0.78 | 1.56 |
|   | AFM-NSO | 0.66 | 1.33 |
|   | AFM-SO  | 0.75 | 1.51 |
| 3 | NSP-NSO | 1.70 | 2.34 |
|   | NSP-SO  | 1.48 | 2.12 |
|   | SP-NSO  | 0.88 | 1.26 |
|   | SP-SO   | 0.99 | 1.40 |
|   | AFM-NSO | 0.85 | 1.22 |
|   | AFM-SO  | 0.98 | 1.43 |
| 4 | NSP-NSO | 1.83 | 2.24 |
|   | NSP-SO  | 1.65 | 2.17 |
|   | SP-NSO  | 0.97 | 1.23 |
|   | SP-SO   | 1.09 | 1.40 |
|   | AFM-NSO | 0.95 | 1.25 |
|   | AFM-SO  | 1.09 | 1.41 |
| 5 | NSP-NSO | 1.91 | 2.24 |
|   | NSP-SO  | 1.73 | 2.03 |
|   | SP-NSO  | 1.04 | 1.29 |
|   | SP-SO   | 1.15 | 1.40 |
|   | AFM-NSO | 1.00 | 1.20 |
|   | AFM-SO  | 1.15 | 1.40 |
| 6 | NSP-NSO | 1.97 | 2.26 |
|   | NSP-SO  | 1.79 | 2.06 |
|   | SP-NSO  | 1.07 | 1.27 |
|   | SP-SO   | 1.20 | 1.43 |
|   | AFM-NSO | 1.04 | 1.24 |
|   | AFM-SO  | 1.19 | 1.37 |
| 7 | NSP-NSO | 2.01 | 2.26 |
|   | NSP-SO  | 1.82 | 2.05 |
|   | SP-NSO  | 1.10 | 1.25 |
|   | SP-SO   | 1.23 | 1.42 |
|   | AFM-NSO | 1.07 | 1.22 |
|   | AFM-SO  | 1.21 | 1.36 |

Table II. Magnetic moments MM per atom ($\mu_B$/atom), spin polarization energies per atom $E_{sp}$, spin orbit coupling energies per atom $E_{so}$, for the Am (001) N layers (N=1-7).

| N | Theory | MM ($\mu_B$/atom) | $E_{sp}$ (eV/atom) | $E_{so}$ (eV/atom) |
|---|---|---|---|---|
| 1 | NSP-NSO |  |  |  |
|   | NSP-SO |  |  | 8.94 |
|   | SP-NSO | 7.70 | 3.49 |  |
|   | SP-SO | 7.52 | 2.11 | 7.56 |
|   | AFM-NSO | 7.81 | 3.53 |  |
|   | AFM-SO | 7.60 | 2.17 | 7.58 |
| 2 | NSP-NSO |  |  |  |
|   | NSP-SO |  |  | 8.73 |
|   | SP-NSO | 7.32 | 2.81 |  |
|   | SP-SO | 7.01 | 1.73 | 7.65 |
|   | AFM-NSO | 0 | 2.82 |  |
|   | AFM-SO | 0 | 1.76 | 7.67 |
| 3 | NSP-NSO |  |  |  |
|   | NSP-SO |  |  | 8.73 |
|   | SP-NSO | 7.33 | 2.68 |  |
|   | SP-SO | 7.02 | 1.62 | 7.67 |
|   | AFM-NSO | 2.43 | 2.69 |  |
|   | AFM-SO | 2.39 | 1.67 | 7.71 |
| 4 | NSP-NSO |  |  |  |
|   | NSP-SO |  |  | 8.76 |
|   | SP-NSO | 7.31 | 2.63 |  |
|   | SP-SO | 6.93 | 1.55 | 7.68 |
|   | AFM-NSO | 0 | 2.65 |  |
|   | AFM-SO | 0 | 1.60 | 7.72 |
| 5 | NSP-NSO |  |  |  |
|   | NSP-SO |  |  | 8.76 |
|   | SP-NSO | 7.35 | 2.61 |  |
|   | SP-SO | 6.97 | 1.53 | 7.68 |
|   | AFM-NSO | 1.46 | 2.62 |  |
|   | AFM-SO | 1.41 | 1.59 | 7.73 |
| 6 | NSP-NSO |  |  |  |
|   | NSP-SO |  |  | 8.75 |
|   | SP-NSO | 7.35 | 2.59 |  |
|   | SP-SO | 6.97 | 1.53 | 7.69 |
|   | AFM-NSO | 0 | 2.60 |  |
|   | AFM-SO | 0 | 1.57 | 7.73 |
| 7 | NSP-NSO |  |  |  |
|   | NSP-SO |  |  | 8.75 |
|   | SP-NSO | 7.32 | 2.58 |  |
|   | SP-SO | 6.95 | 1.52 | 7.69 |
|   | AFM-NSO | 1.03 | 2.58 |  |
|   | AFM-SO | 1.01 | 1.56 | 7.73 |

Table III. Work functions $W$ (in eV) and surface energies (in J/m$^2$) for fcc Am (001) N layers (N=1-7).

| N | Theory | $W$ (eV) | $E_s$ (J/m$^2$) |
|---|---|---|---|
| 1 | NSP-NSO | 3.32 | 2.11 |
| | NSP-SO | 3.15 | 1.76 |
| | SP-NSO | 2.70 | 0.75 |
| | SP-SO | 2.80 | 0.88 |
| | AFM-NSO | 2.77 | 0.80 |
| | AFM-SO | 2.86 | 0.93 |
| 2 | NSP-NSO | 3.37 | 1.65 |
| | NSP-SO | 3.34 | 1.55 |
| | SP-NSO | 2.71 | 0.68 |
| | SP-SO | 2.80 | 0.79 |
| | AFM-NSO | 2.78 | 0.73 |
| | AFM-SO | 2.90 | 0.85 |
| 3 | NSP-NSO | 3.36 | 1.56 |
| | NSP-SO | 3.31 | 1.51 |
| | SP-NSO | 2.69 | 0.68 |
| | SP-SO | 2.85 | 0.79 |
| | AFM-NSO | 2.80 | 0.73 |
| | AFM-SO | 2.91 | 0.82 |
| 4 | NSP-NSO | 3.31 | 1.57 |
| | NSP-SO | 3.28 | 1.42 |
| | SP-NSO | 2.66 | 0.70 |
| | SP-SO | 2.80 | 0.80 |
| | AFM-NSO | 2.78 | 0.72 |
| | AFM-SO | 2.96 | 0.81 |
| 5 | NSP-NSO | 3.33 | 1.58 |
| | NSP-SO | 3.31 | 1.46 |
| | SP-NSO | 2.68 | 0.69 |
| | SP-SO | 2.82 | 0.81 |
| | AFM-NSO | 2.79 | 0.73 |
| | AFM-SO | 2.89 | 0.81 |
| 6 | NSP-NSO | 3.36 | 1.57 |
| | NSP-SO | 3.27 | 1.47 |
| | SP-NSO | 2.67 | 0.68 |
| | SP-SO | 2.81 | 0.80 |
| | AFM-NSO | 2.80 | 0.73 |
| | AFM-SO | 2.96 | 0.82 |
| 7 | NSP-NSO | 3.30 | 1.56 |
| | NSP-SO | 3.28 | 1.48 |
| | SP-NSO | 2.67 | 0.69 |
| | SP-SO | 2.78 | 0.80 |
| | AFM-NSO | 2.78 | 0.73 |
| | AFM-SO | 2.93 | 0.82 |

## List of figures

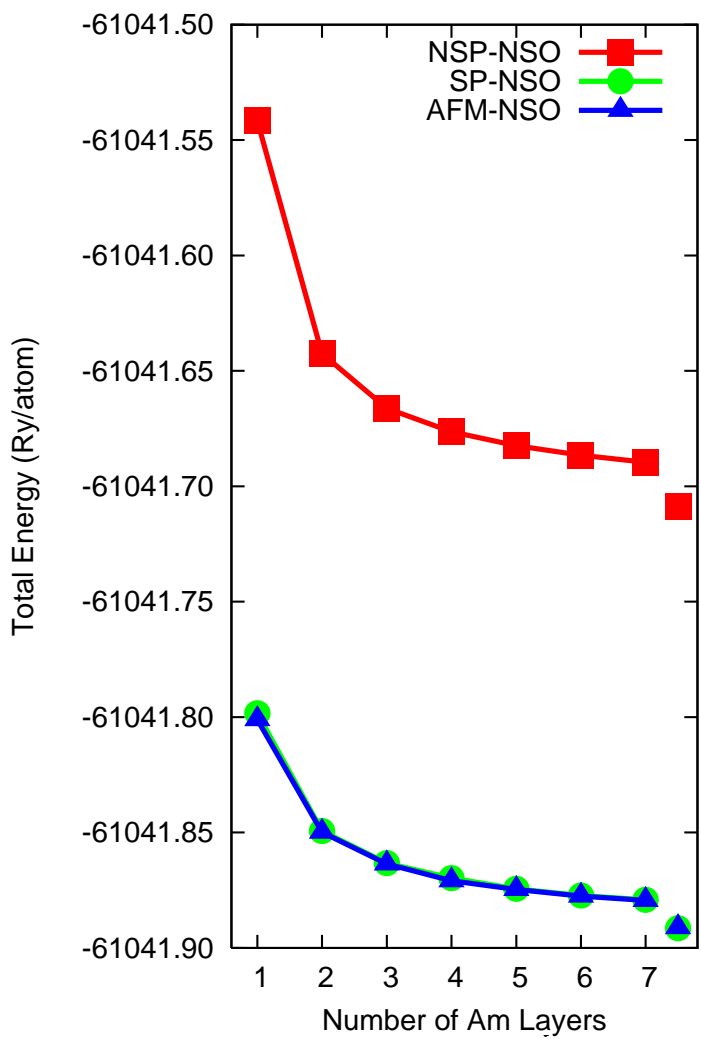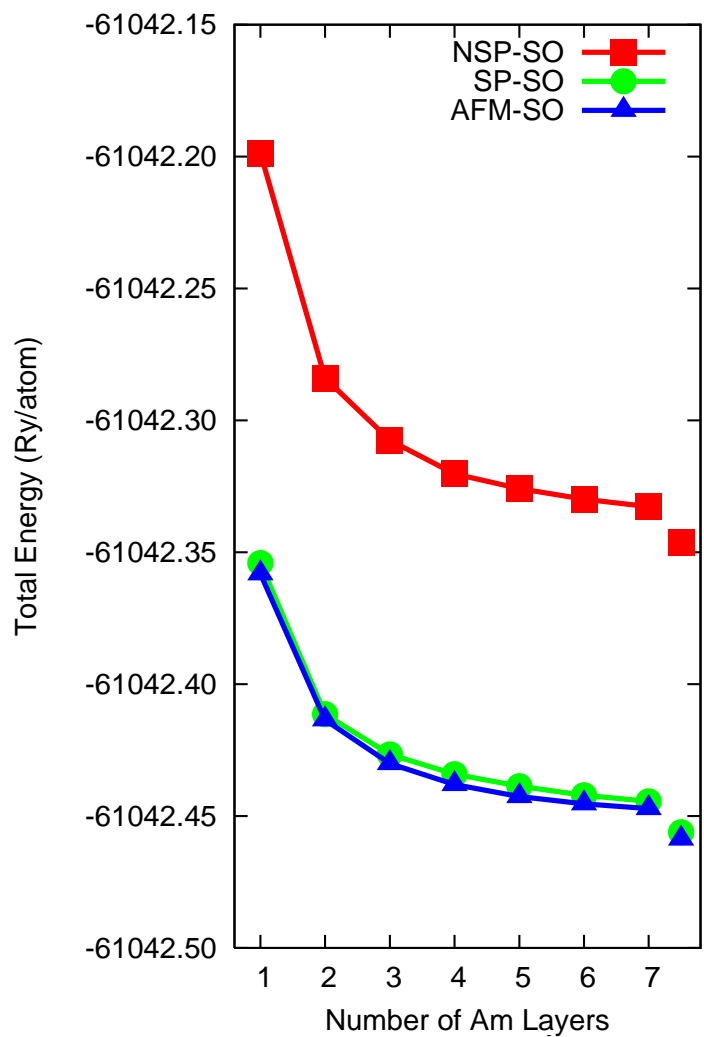

Fig. 1. Total energies of Am (001) films for different number of layers (N = 1 - 7). For comparison, the corresponding total energies of Am bulk are also plotted at the farthest right of the figure.

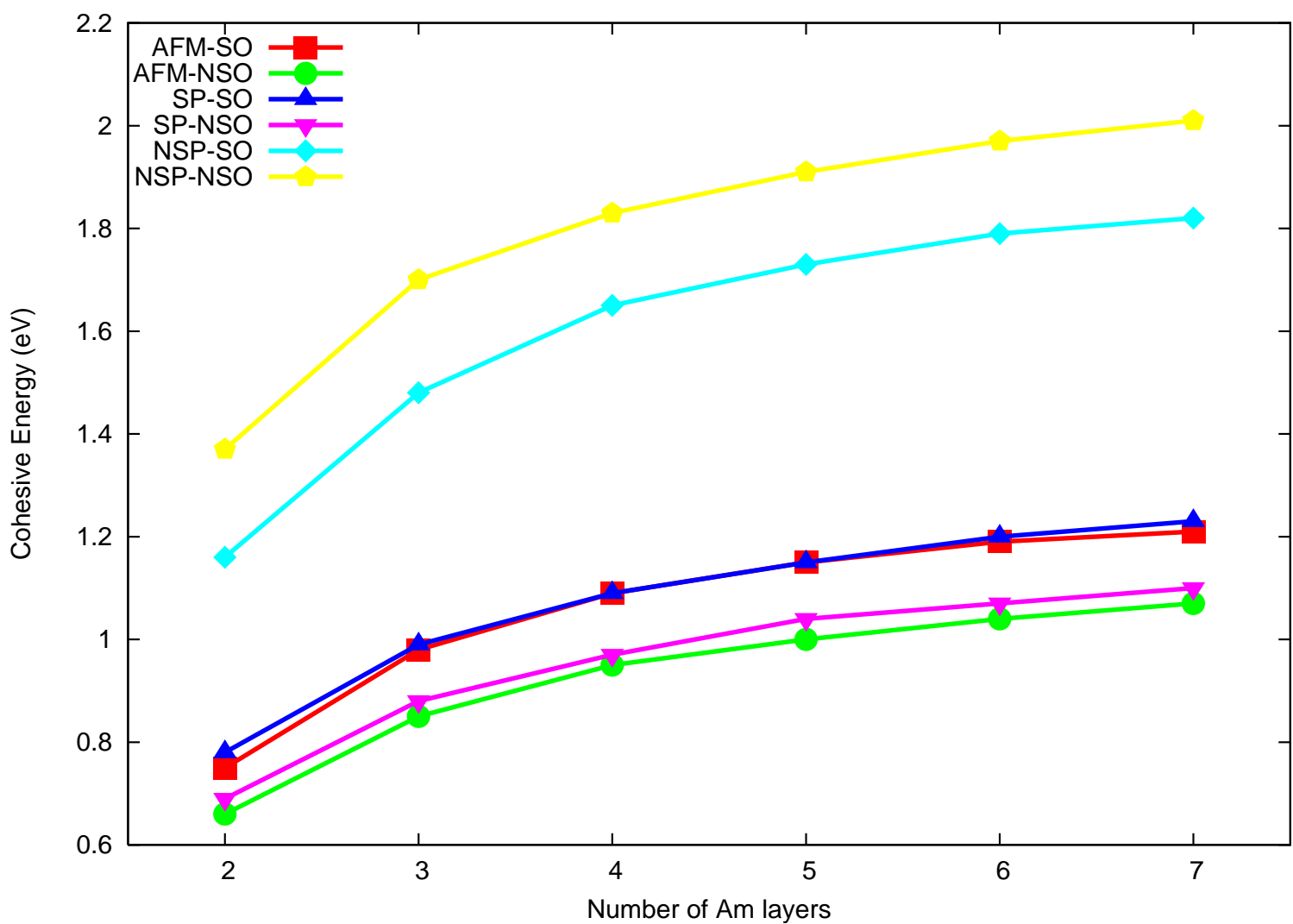

Fig. 2. Cohesive energies per atom of the fcc Am (001) films with respect to the Am monolayer versus the number of Am layers.

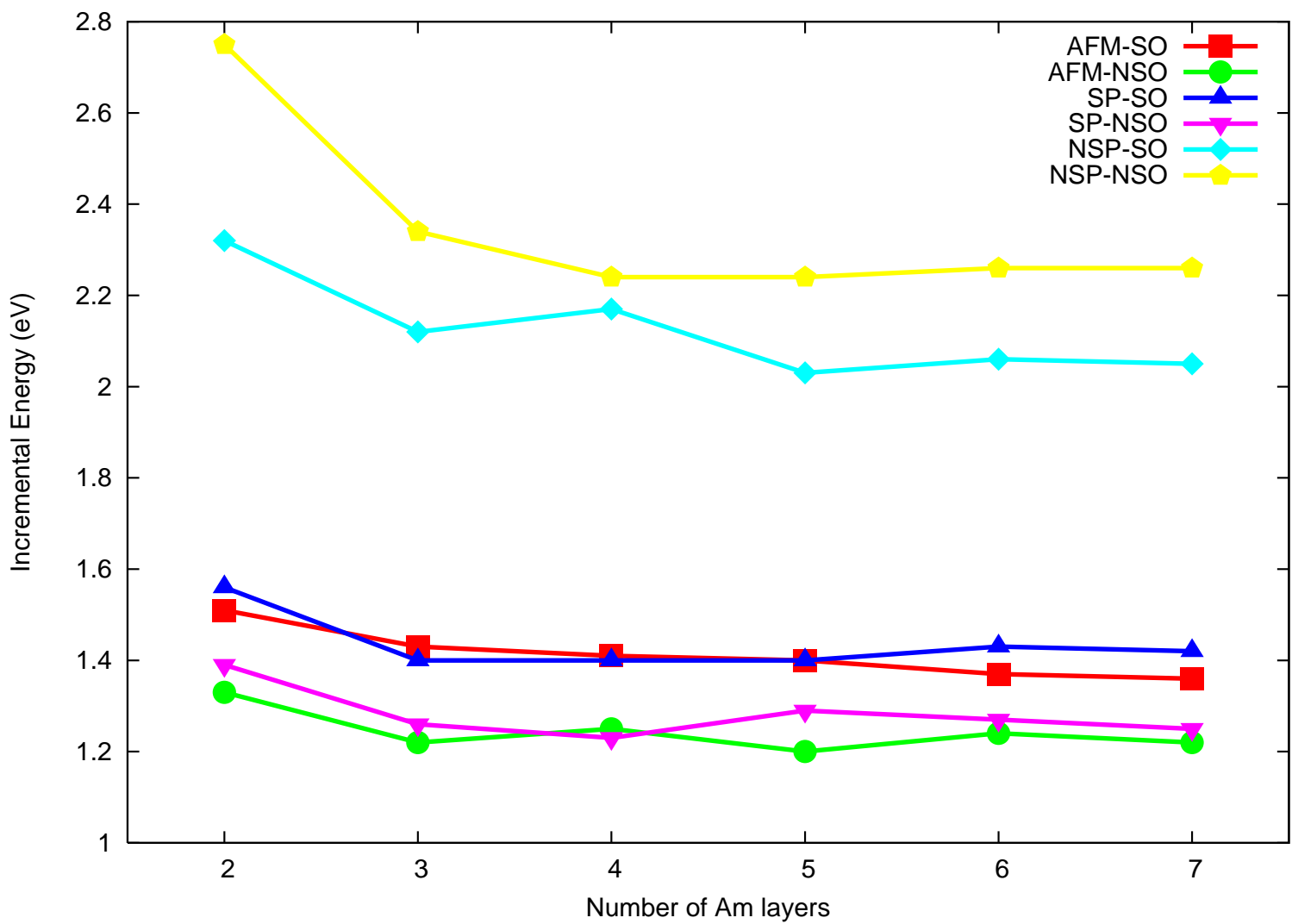

Fig. 3. Incremental energies per atom of the fcc Am (001) films with respect to the (N-1) fcc Am films plus the monolayer versus the number of Am layers

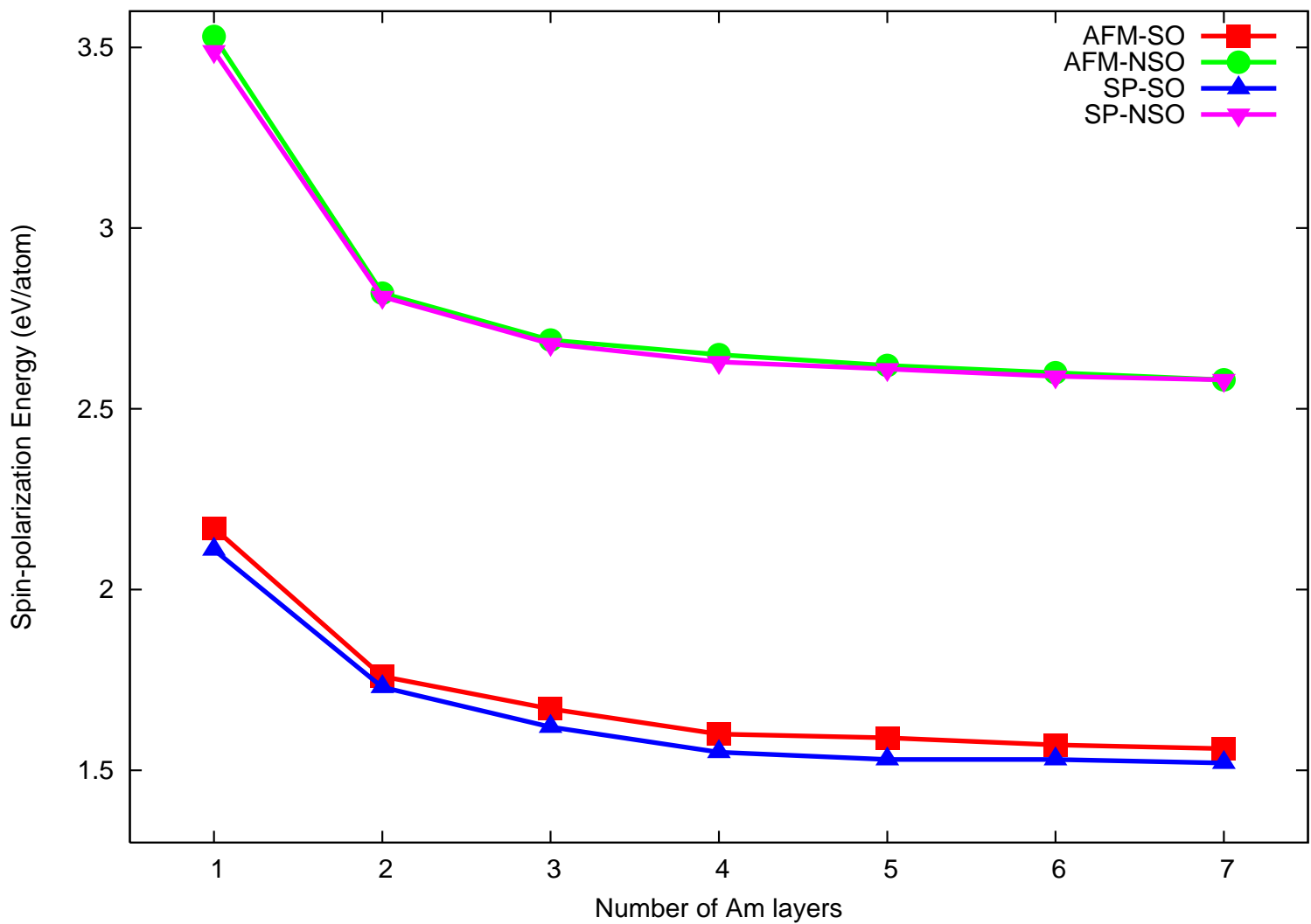

Fig. 4. Spin-polarization energies (eV/atom) as a function of the number of Am layers for Am (001) films for N layers (N=1-7).

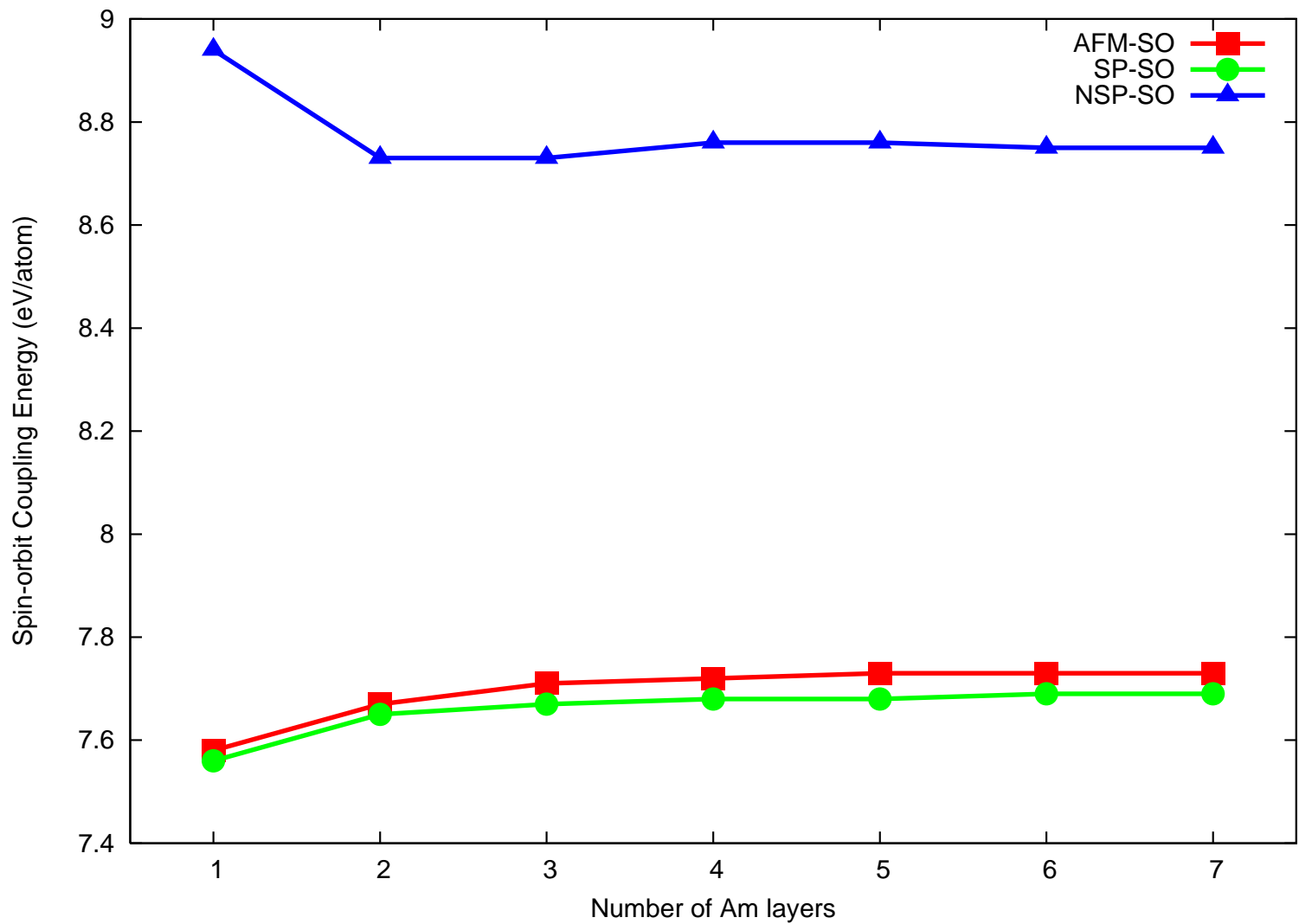

Fig. 5. Spin-orbit coupling energies (eV/atom) as a function of the number of Am layers for Am (001) films for N layers (N=1-7).

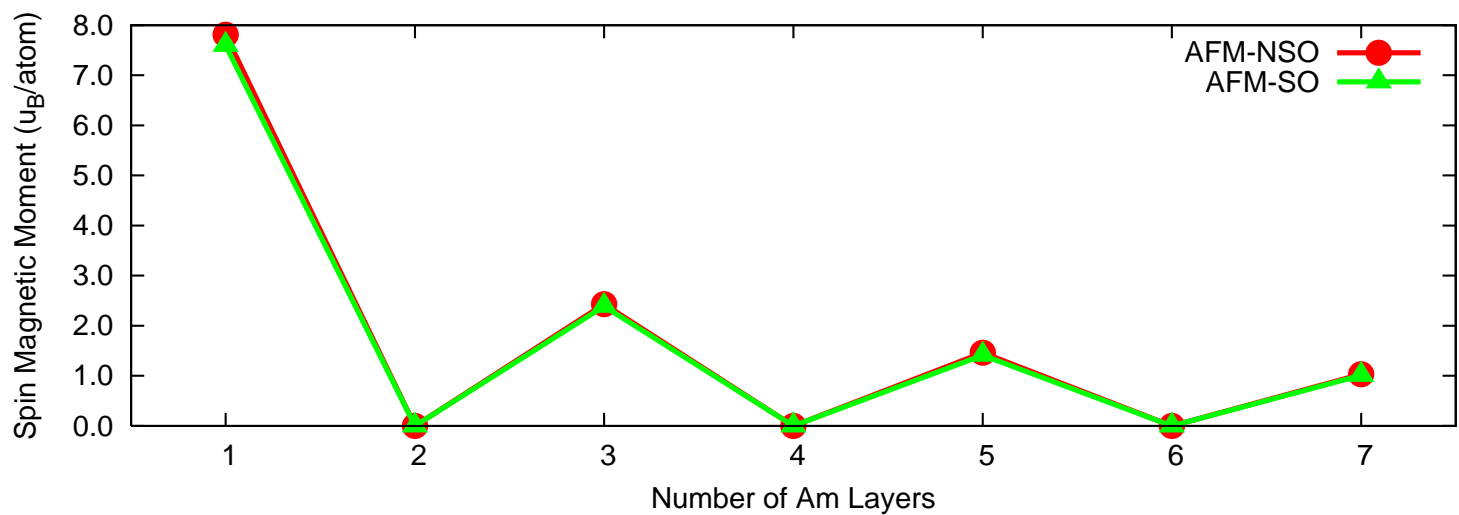
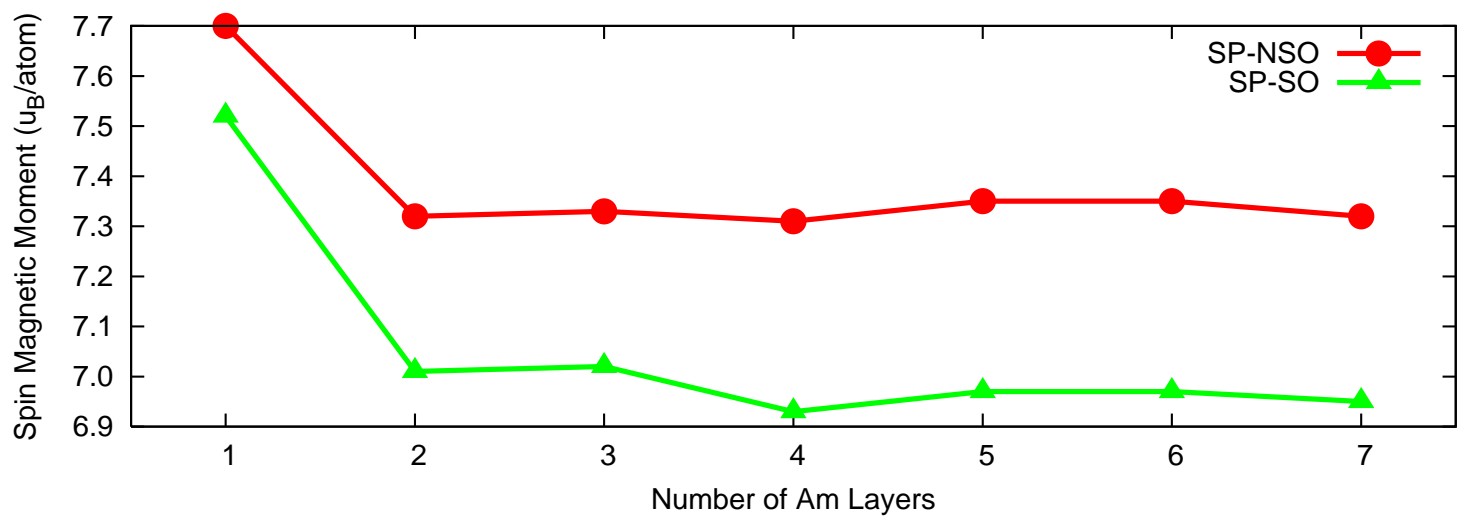

Fig.6. Spin magnetic moments of the fcc Am (001) films for different layers (N=1-7).

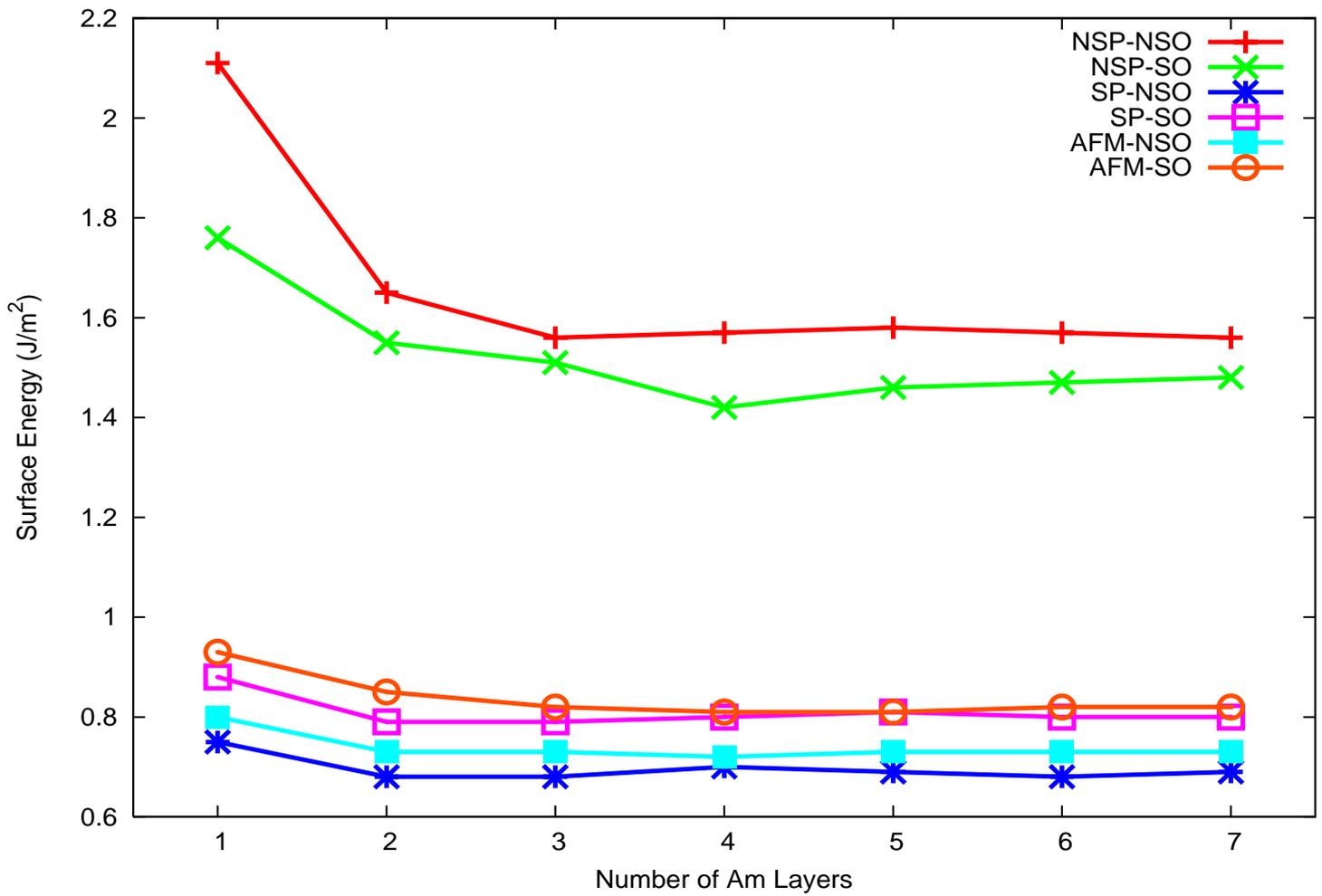

Fig. 7. Surface energies for the fcc Am (001) films vs. the number of Am layers.

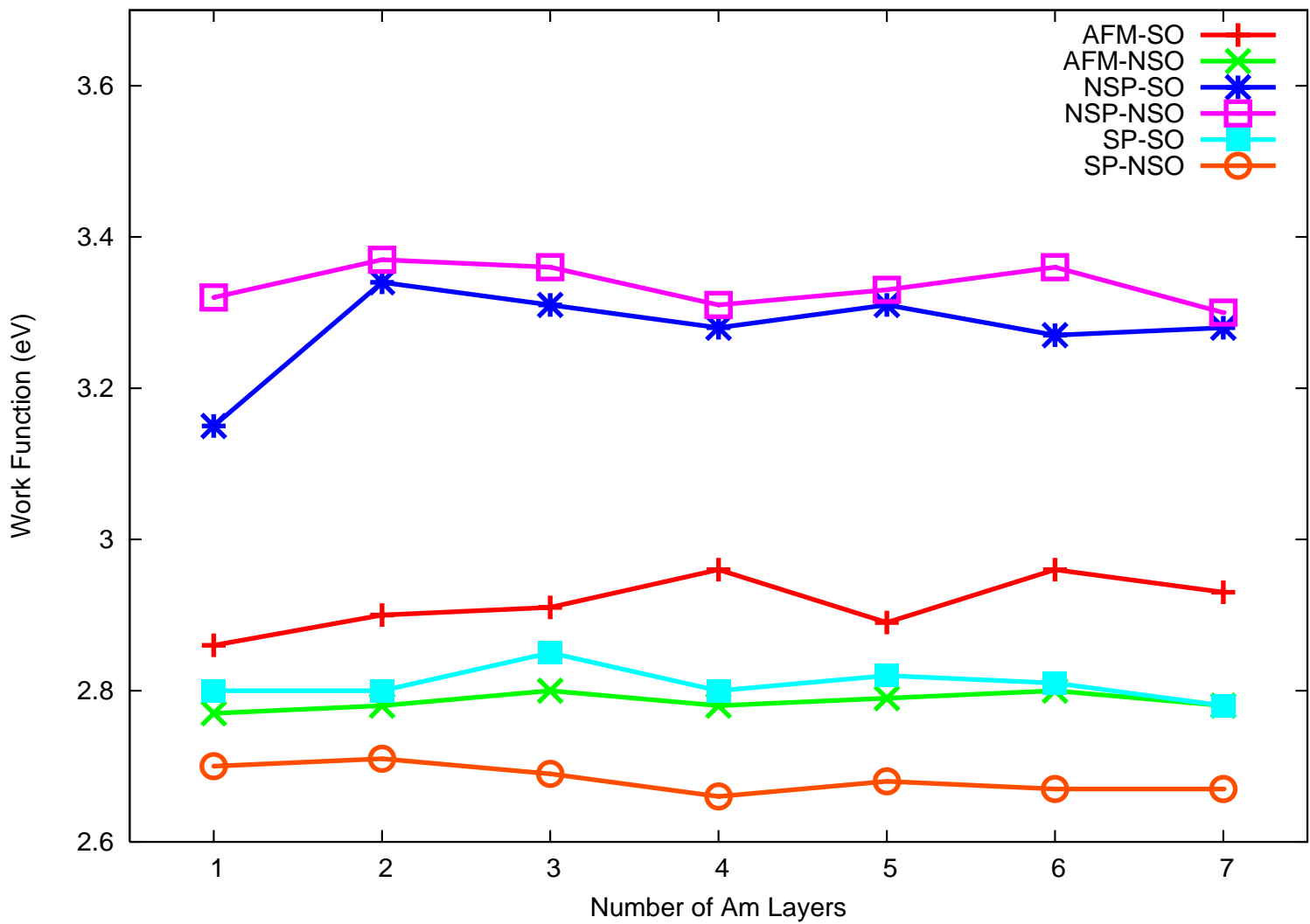

Fig. 8. Work functions of the fcc Am films (eV) vs. the number of Am (001) layers.

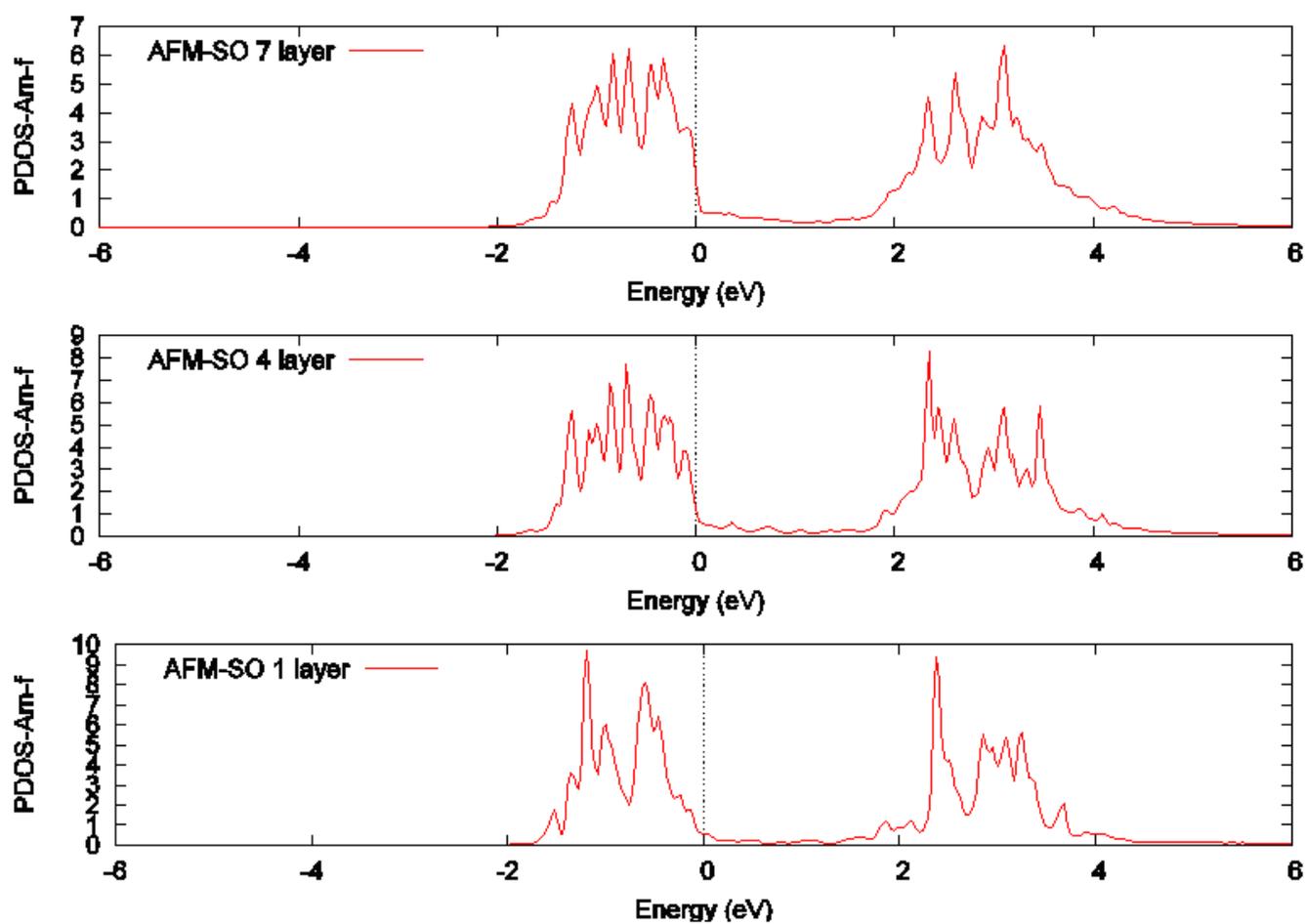

Fig. 9. Density of states of 5f electrons for Am (001) N-layer films at the ground state, where N = 1, 4, 7 as labeled in the figure. Fermi energy is set at zero.